\input harvmac
\input epsf

\def\eps2{\epsilon_2}


\Title{ \vbox{\baselineskip12pt \rightline{HUTP-03/A001}
 \rightline{hep-th/0301029}
 \vskip -.25in} }{ \vbox{
\centerline{On Curvature-Squared Corrections for}
\vskip 0.1in
\centerline{ D-brane Actions} } }

\centerline{ Martijn Wijnholt}
\vskip .1in
\centerline{\sl Jefferson Laboratory of Physics, Harvard
University}
\centerline{\sl Cambridge, MA 02138, USA}
\centerline{\it wijnholt@fas.harvard.edu }

\vskip .1in

\vskip 0.3in

\centerline{\bf Abstract}

\noindent
Curvature squared corrections for D-brane actions in type II string theory
were derived by Bachas, Bain and Green. Here we write down a
generalisation of these corrections to all orders in $F$, the field
strength of the U(1) gauge field on the brane.  Some of these terms
are needed to restore consistency with T-duality.

\Date{January 2003}


\lref\TseytlinDJ{
A.~A.~Tseytlin,
``Born-Infeld action, supersymmetry and string theory,''
arXiv:hep-th/9908105.
}

\lref\AbouelsaoodGD{
A.~Abouelsaood, C.~G.~Callan, C.~R.~Nappi and S.~A.~Yost,
``Open Strings In Background Gauge Fields,''
Nucl.\ Phys.\ B {\bf 280}, 599 (1987).
}

\lref\GarousiAD{
M.~R.~Garousi and R.~C.~Myers,
``Superstring Scattering from D-Branes,''
Nucl.\ Phys.\ B {\bf 475}, 193 (1996)
[arXiv:hep-th/9603194].
}

\lref\BachasUM{
C.~P.~Bachas, P.~Bain and M.~B.~Green,
``Curvature terms in D-brane actions and their M-theory origin,''
JHEP {\bf 9905}, 011 (1999)
[arXiv:hep-th/9903210].
}

\lref\JohnsonQT{
C.~V.~Johnson, A.~W.~Peet and J.~Polchinski,
``Gauge theory and the excision of repulson singularities,''
Phys.\ Rev.\ D {\bf 61}, 086001 (2000)
[arXiv:hep-th/9911161].
}

\lref\WijnholtUS{
M.~Wijnholt and S.~Zhukov,
``Inside an enhancon: Monopoles and dual Yang-Mills theory,''
Nucl.\ Phys.\ B {\bf 639}, 343 (2002)
[arXiv:hep-th/0110109].
}

\lref\GarousiBJ{
M.~R.~Garousi,
``Superstring scattering from D-branes bound states,''
JHEP {\bf 9812}, 008 (1998)
[arXiv:hep-th/9805078].
}

\lref\ArdalanQT{
F.~Ardalan, H.~Arfaei, M.~R.~Garousi and A.~Ghodsi,
``Gravity on noncommutative D-branes,''
arXiv:hep-th/0204117.
}

\lref\FotopoulosPT{
A.~Fotopoulos,
``On (alpha')**2 corrections to the D-brane action for non-geodesic  world-volume embeddings,''
JHEP {\bf 0109}, 005 (2001)
[arXiv:hep-th/0104146].
}

\lref\GarousiAD{
M.~R.~Garousi and R.~C.~Myers,
``Superstring Scattering from D-Branes,''
Nucl.\ Phys.\ B {\bf 475}, 193 (1996)
[arXiv:hep-th/9603194].
\hfill \break
I.~R.~Klebanov and L.~Thorlacius,
``The Size of p-Branes,''
Phys.\ Lett.\ B {\bf 371}, 51 (1996)
[arXiv:hep-th/9510200].
\hfill \break
A.~Hashimoto and I.~R.~Klebanov,
``Scattering of strings from D-branes,''
Nucl.\ Phys.\ Proc.\ Suppl.\  {\bf 55B}, 118 (1997)
[arXiv:hep-th/9611214].
}

\lref\PolchinskiRQ{
J.~Polchinski,
``String Theory.'' Vols. 1 \& 2.
}

\lref\WyllardYE{
N.~Wyllard,
``Derivative corrections to the D-brane Born-Infeld action: Non-geodesic  embeddings and the Seiberg-Witten map,''
JHEP {\bf 0108}, 027 (2001)
[arXiv:hep-th/0107185].
\hfill \break
N.~Wyllard,
``Derivative corrections to D-brane actions with constant background  fields,''
Nucl.\ Phys.\ B {\bf 598}, 247 (2001)
[arXiv:hep-th/0008125].
}

%

The low energy effective action for D-branes carries interesting
information about the physics of D-branes as well as the space-time it lives in.
The massless fields usually include a Yang-Mills multiplet, and the
action for these fields to lowest order is then given by the standard
Yang-Mills action. Many higher order corrections have been found, and in
general we should expect every possible correction to be present unless it
is forbidden by symmetries (or shown not to exist by calculation).
There are however some corrections that can be calculated to all orders
with relative ease and that carry interesting information about D-brane
physics. The prime example of these are the $F^n$ corrections where $F$ is
the field strength of the worldvolume gauge field. At tree level,
these sum into the (Dirac)-Born-Infeld
action, as has been established by several different arguments.\foot{
For some history and references, see \TseytlinDJ.}

In this note we would like to write down tree level
 corrections for D-branes in type II string theory of the form $R^2 F^n$
to all orders in $F$. Here $R$ denotes the pull-back of the
space-time curvature and we are referring to terms in the CP-even part
 of the action. There are similar terms in the Chern-Simons part of
 the action which are well known. In the limit $F \to 0$ we reduce to the
curvature-squared corrections found in \BachasUM\ (see also \FotopoulosPT).
Since other corrections in $F$ to curvature-squared terms will involve  derivatives, these
 terms can still be trusted for uniform near-critical field strengths.

While there may be several reasons for wanting to know such
corrections,
our interest arose from an apparent
inconsistency with T-duality. When a D-brane is wrapped on a K3
surface, the curvature--squared term produces a correction to the
tension of the brane after integration over the K3. In \JohnsonQT\
the known terms in the Lagrangian\foot{
We will be schematic here, more precise expressions appear
below. Also, a factor of $2\pi \alpha'$ will be absorbed in $F$.}
\eqn\aa{ e^{-\varphi} \sqrt{{\rm det}(1+F)} + e^{-\varphi} R^2
}
were used to get relations between the
worldvolume gauge theory and paramaters in the supergravity
solution for such wrapped branes. It was pointed out in \WijnholtUS\ that these formulae
are not consistent with the $SO(20,4;Z)$ T-duality group of the
K3. The resolution of this puzzle was also mentioned in
\WijnholtUS: in the presence of non-constant $F$, the $R^2$ corrections
to the action should take the form
\eqn\corBI{
 e^{-\varphi} R^2 \to e^{-\varphi} \sqrt{{\rm det}(1 + F)}\, R^2 + \ldots .}
It is not too hard to verify the presence of such terms; the factor of
$\sqrt{{\rm det}(1 + F)}$ in front of $R^2$ comes from the partition
function on the disk
when there is a constant $F$ background. The ellipses stand for
corrections involving $R$ and $F$ with cross-contractions.
Now when a D-brane is wrapped on K3 with volume $V$ and the $R^2$ term is
integrated, the reduced action is of the form
\eqn\aa{ \left( {V\over (2\pi \sqrt{\alpha'})^4} - 1 \right)
\int e^{-\varphi} \sqrt{{\rm det}(1 + F)} }
and this restores consistency with T-duality. Thus the $R^2$ term,
besides correcting the tension,
also corrects a host of other couplings, such as the effective Yang-Mills coupling on the
unwrapped part of the D-brane.

In the remainder we would like to try to fix the remaining $R^2$
corrections which involve cross-contractions with $F$ in \corBI.
We will do this by trying to write down vertices that reproduce
the string amplitude for the scattering of a graviton off a D-brane
to order $\alpha'^2$.

The string amplitude that we will use  is for the scattering of a graviton
from a D-brane with constant electric-magnetic field on it. It was
first written down in \GarousiBJ\foot{
For the similar computation of graviton scattering with $F=0$, see \GarousiAD.}
but we have checked it independently.
In order to write down the string amplitude, we need to introduce some
notation. The graviton polarisation tensors will be denoted by $\epsilon_i$
and the graviton momenta by $p_i$. The polarisation tensors are symmetric
and on-shell
\eqn\kin{ \eqalign{ &\epsilon_{i\, \mu\nu} = \epsilon_{i\, \nu\mu}, \
 \epsilon_{i\, \mu}^{\ \ \mu} = 0 ,\cr
  &\epsilon_{i\, \mu\nu}p_i^\mu = \epsilon_{i\, \mu\nu}p_i^\nu = 0 \ \forall i.}}
Similarly the momenta satisfy $p_i^2 = 0$. This is compatible with turning
on a constant $F$, since the curvature of the field $B + F$ is still identically
zero and does not source the bulk  Ricci tensor. In this paper we will also assume that
the second fundamental form vanishes identically. We will comment on relaxing this
assumption later.

We are interested in computing
a disk amplitude with two graviton vertex operators
inserted in the interior of the disk and an arbitrary number of photon
vertex operators on the boundary of the disk.  The
boundary of the disk is restricted to lie along the worldvolume of a
Dp-brane,  and so are the photon
momenta and polarisations. The polarisation tensors and momenta of the
gravitons  however have no such restrictions.
%
%
\bigskip
\centerline{\epsfxsize=.91\hsize\epsfbox{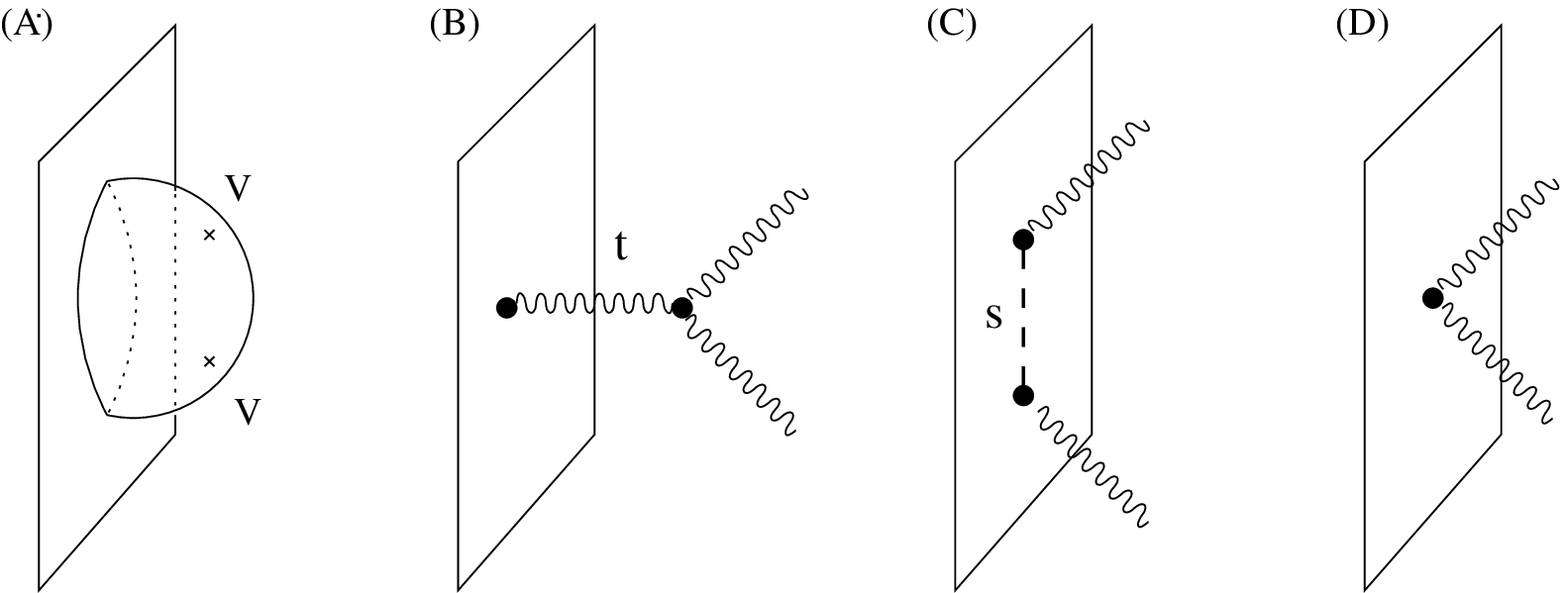}}
\nobreak\noindent{\ninepoint\sl \baselineskip=8pt \fig\disk{}:
{\sl (A): Relevant disk diagram. The following pictures arise in the
low energy limit. (B): t-channel diagram. (C): s-channel diagram. (D):
Contact interactions.}}
\bigskip


In principle one would need to compute infinitely many disk diagrams, with
various numbers of photon vertex operators inserted on the boundary.
However since we are assuming the
photon field strength  to be uniform, we may
effectively sum over all the contractions between the gravitons and the
photons at once by introducing an $F$ dependent propagator  \AbouelsaoodGD.
So we only need a single disk diagram with two graviton insertions in the
interior. Let us put $\alpha' = 2$ in the next few equations.
The amplitude is of the form
\eqn\aa{ {\cal A} \sim \int d^2z_1 d^2z_2
\bigl< V_1(z_1,\bar{z}_1) V_2(z_2,\bar{z}_2) \bigr> .
}
We will take the first graviton vertex operator in the (-1,-1) picture and
the second in the (0,0) picture:
\eqn\aa{ \eqalign{
V_1(z_1,\bar{z}_1)&= \epsilon_{1\mu\nu}
:e^{-\phi(z_1)} \psi^\mu(z_1) e^{i p_1\cdot X(z_1)} :
:e^{-\tilde{\phi}(\bar{z}_1)} \bar{\psi}^\nu(\bar{z}_1) e^{i p_1\cdot
\tilde{X}(\bar{z}_1)}  : \cr
V_2(z_2,\bar{z}_2) &= \epsilon_{2\rho\sigma}
:(\del X^\rho + i p_2 \cdot \psi\ \psi^\rho) e^{i p_2 \cdot
X} :
:(\bar{\del} \bar{X}^\sigma + i p_2 \cdot \bar{\psi}\
\bar{\psi}^\sigma) e^{i p_2 \cdot
\tilde{X}} : }}
The contractions between leftmovers are the usual ones:
\eqn\aa{ \eqalign{
\left< X(z)^\mu {X}^\nu ({z}')\right> &= - \eta^{\mu\nu} \log(z-{z}') \cr
\bigl< \psi^\mu(z) \psi^\nu(z') \bigr> &= -{\eta^{\mu\nu}\over z-z'} \cr
\bigl< \phi(z) \phi(z') \bigr> &= - \log (z - z') }}
and similarly for the rightmovers. The contractions between left- and
rightmovers with indices will be modified due to the Dirichlet and
$F$-dependent boundary
conditions on the propagators, so we will have
\eqn\aa{\eqalign{
\bigl< X(z)^\mu \tilde{X}^\nu(\bar{z}')\bigr> &=
-D_F^{\mu\nu} \log(z-\bar{z}') \cr
\bigl< \psi^\mu(z) \bar{\psi}^\nu(\bar{z}') \bigr> &= - {D_F^{\mu\nu}\over z
- \bar{z}'} \cr
\bigl< \phi(z) \tilde{\phi}(\bar{z}') \bigr> &= - \log(z - \bar{z}') }}
where
\eqn\aa{
D(F)_{\mu\nu} = \left\{ \matrix{ \left({1-F \over 1+F}\right)^{\mu\nu}
{\rm for\ tangent \ indices} \cr
        - \delta^{\mu\nu} \qquad {\rm \ for\ normal \ indices.} }\right.  }
Notice that $D(F)^T = D(F)^{-1}= D(-F)$, i.e. $D(F)$ is an orthogonal matrix.

Let us also introduce the kinematic variables  $p_1 \cdot p_2 = -t/2$ where $t$ is
interpreted as the momentum transfer between the incoming graviton and the
brane, and  $p_1 \cdot D_F \cdot p_1 = p_2 \cdot D_F \cdot p_2 = 2
q_F^2 \equiv - 2 s$.
 Then one finds the following amplitude (with a slight rearranging compared to \GarousiBJ):
\eqn\stringamp{
{\cal A} \sim   T_p  {\Gamma(2 q_F^2) \Gamma(-t/2)\over \Gamma(1 + 2 q_F^2 -
t/2)} \sqrt{ {\rm det}(\eta_{\mu\nu} + F_{\mu\nu})}
\left( 2 q_F^2 a^F_1 - {t \over 2}\,  a^F_2 \right)  .}
with
\eqn\aa{\eqalign{
a_1^F =&- {\rm Tr}(\epsilon_1 \cdot D_F)(p_1 \cdot \eps2 \cdot p_1)
- {\rm Tr}(\eps2 \cdot D_F)(p_2 \cdot \epsilon_1 \cdot p_2)
-2 q_F^2 {\rm Tr}(\epsilon_1 \cdot \eps2) \cr
& + (p_2 \cdot \epsilon_1 \cdot D_F \cdot \eps2 \cdot p_1)
+ (p_2 \cdot \epsilon_1 \cdot D_F^T \cdot \eps2 \cdot p_1) \cr
&+  (p_2 \cdot D_F \cdot \epsilon_1 \cdot \eps2 \cdot D_F^T \cdot p_1)
+  (p_2 \cdot D_F^T \cdot \epsilon_1 \cdot \eps2 \cdot D_F \cdot
p_1)\cr
&- (p_1 \cdot D_F \cdot \epsilon_1 \cdot \eps2 \cdot D_F^T \cdot
p_2)
- (p_1 \cdot D_F^T \cdot \epsilon_1 \cdot \eps2 \cdot D_F \cdot p_2)  }}
and
\eqn\aa{\eqalign{
a_2^F=& -{\rm Tr}(\epsilon_1 \cdot D_F)(p_1 \cdot \eps2 \cdot p_1)
-{\rm Tr}(\eps2 \cdot D_F)(p_2 \cdot \epsilon_1 \cdot
p_2)\cr
&+ {\rm Tr}(\epsilon_1 \cdot D_F)(p_1 \cdot D_F \cdot \eps2 \cdot D_F \cdot p_1)
+{\rm Tr}(\eps2 \cdot D_F)(p_2 \cdot D_F \cdot \epsilon_1 \cdot D_F \cdot
p_2) \cr
&-2 q_F^2  {\rm Tr}(\epsilon_1 \cdot \eps2)-(2 q_F^2 - t/2){\rm Tr}(\epsilon_1 \cdot D) {\rm Tr}(\eps2\cdot D)
+ 2 q_F^2 {\rm Tr}(\epsilon_1 \cdot D_F \cdot \eps2 \cdot D_F ) \cr
&+ (p_1 \cdot D_F^T \cdot \epsilon_1 \cdot D_F^T \cdot \eps2 \cdot D_F^T \cdot p_2)
+ (p_1 \cdot D_F \cdot \epsilon_1 \cdot D_F \cdot \eps2 \cdot D_F \cdot
p_2) \cr
& - (p_1 \cdot D_F \cdot \epsilon_1 \cdot \eps2 \cdot D_F^T \cdot p_2)
- (p_1 \cdot D_F^T \cdot \epsilon_1 \cdot \eps2 \cdot D_F \cdot p_2) .}}
$T_p$ is the D-brane tension, excluding the factor of $1/g_c$.
The amplitude is invariant under the replacement $F \to -F$. The factor of
$\sqrt{ {\rm det}(\eta_{\mu\nu} + F_{\mu\nu})}$ comes from the partition
function on the disk with $F$-dependent boundary conditions, which
multiplies the amplitude.  Notice that
the positions of the open string poles are changed when $F \not = 0$. This
indicates a change in the effective tension of an open string depending
on direction. Since $F$ is uniform and translation symmetry is unbroken along the brane,
linear momentum in the worldvolume directions is conserved, and one makes heavy use of this
as well as the identities \kin\ in the course of the derivation.

Next we turn to effective actions. In the low energy limit we can expand the string amplitude,
with $\alpha'$ restored,  as
\eqn\expand{ {\cal A} \sim   T_p \sqrt{ {\rm det}(\eta_{\mu\nu} + F_{\mu\nu})}
\left( -2 s\,  a^F_1 - {t \over 2}\,  a^F_2 \right)
\left({\alpha'\over 2}\right)^2
\left( {1\over \alpha'^2 s t} - {\pi^2 \over 6} + {\cal O}(\alpha') \right). }
Now the leading term in this string amplitude
can be reproduced by the usual Dirac-Born-Infeld action plus the 
$\sqrt{g}\, R$  bulk supergravity
Lagrangian.\foot{
By leading order here we mean we have included all order contributions in $F$
 but treat any additional corrections as higher order. For this reason we still keep a factor of
 $2\pi \alpha'$ absorbed in $F$.}
This was essentially shown in \ArdalanQT; this paper deals with the bosonic case,
but the leading order terms in the string amplitudes
(equations (22)-(24) in this paper) coincide with the
superstring and so the calculation carries over.
The next order contribution for the superstring has two extra powers of $\alpha'$.
Here we expect no t-channel
contributions from the effective action, since the relevant 3-point vertices in the bulk
receive no higher order corrections (as one can check from the string
amplitude for three massless NS bosons \PolchinskiRQ). Similarly one
can show there are no new contributions from the s-channel.
As in the computations for the lowest order
contributions to graviton  scattering, all order contributions in $F$ to graviton-ripple
mixing as well as propagators on the brane
 can be grouped together and described by the standard
DBI action, so that the claim becomes that there are no new
contributions beyond these. We can check this by computing a disk diagram with one
graviton and one massless open string vertex operator in a constant $F$ background. The
amplitude coincides
with the leading order terms of the bosonic computation in equation
(66) of \ArdalanQT\ while the higher order terms are absent,
and so we can again use the
calculations of this paper to show that there are indeed no new contributions from the s-channel.
Therefore all the contributions to the subleading term in
\expand\ should
be given by contact interactions. The usual DBI action
does not contain the needed interactions at order $\alpha'^2$, so we are
going to have to add them.

There is a vast amount of available interactions one could add. To
try to simplify the task somewhat, let us consider the limit $F
\to 0$. In this case the new interactions were first derived in
\BachasUM:
\eqn\aa{ \Delta {\cal L} \sim
( R_{\alpha \beta \gamma \delta} R^{\alpha \beta \gamma\delta} -
 2 {\hat R}_{\alpha \beta} {\hat R}^{\alpha \beta}
- R_{a b \alpha\beta} R^{a b \alpha \beta}
+ 2 {\hat R}_{ab} {\hat R}^{ab}) . }
The notation in this expression from \BachasUM\ is: lower alphabet
Greek letters for tangent indices,
Roman letters for normal indices and a hat for contractions over the tangent indices
while excluding the normal indices.
Using vanishing of the Ricci tensor this may be rewritten in the following form:
\eqn\gbb{\eqalign{
  R_{\alpha \beta \gamma \delta} R^{\alpha \beta \gamma\delta} 
- R_{a b \alpha\beta} R^{a b \alpha \beta} &=
{1\over 8} 
R_{\mu\nu\kappa\lambda} R_{\pi\rho\sigma\tau}g^{\mu\pi}D^{\nu\rho}
(g + D)^{\kappa\sigma} (g + D)^{\lambda\tau} \cr
 -2 {\hat R}_{\alpha \beta} {\hat R}^{\alpha \beta}
+ 2 {\hat R}_{ab} {\hat R}^{ab}
&= -{2\over 8} R_{\mu\nu\kappa\lambda}
R_{\pi\rho\sigma\tau}g^{\mu\pi}D^{\kappa\sigma}
D^{\nu\lambda} D^{\rho\tau} .}}
By $D$ we mean $D(F=0)$.
This suggests one might be able to build the new interactions out
of similar looking tensors by replacing $D$ by $D_F$ or $D_F^T$
and writing down all possible orderings. Since the string amplitude is symmetric under
$F \to -F$ of course we want to keep the contact terms symmetric too.
This yields the following
combinations:
\eqn\short{\eqalign{
 R^2_{(1)} &=
 \, R_{abcd}\,  R_{efgh}\,  g^{ae}\,  D_F^{bd}\,  D_F^{fh}\, D_F^{cg}
+   \quad \{ F \to -F\} \cr
 R^2_{(2)} &=
 \, R_{abcd}\,  R_{efgh}\,  g^{ae}\,  D_F^{bd}\, D_F^{fh}\, D_F^{T\, cg}
+   \quad \{ F \to -F\} \cr
 R^2_{(3)} &=
 \, R_{abcd}\,  R_{efgh}\,  g^{ae}\,  D_F^{T\, bd}\, D_F^{fh}\, D_F^{cg}
+   \quad \{ F \to -F\} \cr
 R^2_{(4)} &=
 \, R_{abcd}\,  R_{efgh}\,  g^{ae}\,  D_F^{bd}\, D_F^{T\, fh}\, D_F^{cg}
+   \quad \{ F \to -F\} \cr
 R^2_{(5)} &=
 R_{abcd} R_{efgh} g^{ae} D_F^{ bf} (g + D_F)^{cg} (g + D_F)^{dh}
+ \{ F \to -F\} \cr
 R^2_{(6)} &=
 R_{abcd} R_{efgh} g^{ae} D_F^{T\, bf} (g + D_F)^{cg} (g + D_F)^{dh}
+ \{ F \to -F\} \cr
R^2_{(7)} &=
 R_{abcd} R_{efgh} g^{ae} (g + D_F)^{bf} D_F^{cg} (g + D_F^T)^{dh}
+ \{ F \to -F\} \cr
R^2_{(8)} &=
 R_{abcd} R_{efgh} g^{ae} (g + D_F^T)^{bf} D_F^{cg} (g + D_F)^{dh}
+ \{ F \to -F\} \cr
R^2_{(9)} &=
 R_{abcd} R_{efgh} g^{ae} D_F^{bf} (g + D_F)^{cg} (g + D_F^T)^{dh}
+ \{ F \to -F\} \cr
R^2_{(10)} &=
 R_{abcd} R_{efgh} g^{ae} (g + D_F)^{bf} D_F^{cg} (g + D_F)^{dh}
+ \{ F \to -F\} \cr
R^2_{(11)} &=
 R_{abcd} R_{efgh} g^{ae} (g + D_F^T)^{bf} D_F^{cg} (g + D_F^T)^{dh}
+ \{ F \to -F\}.}}
There are many ways to rewrite the above expressions, but we can
always ask for the first index on each curvature tensor to be contracted
with the metric. It turns out these expressions contain all the
needed contact interactions. Notice that we included some orderings that do not reduce to
either of the terms in \gbb; however one cannot reproduce the string
amplitude without them.
Leaving out the horrendous computation and simply quoting the answer,
we find that the effective Lagrangian is given by
\eqn\result{\eqalign{
{\cal L}  &=
 T_{p}\, e^{-\varphi}  \sqrt{ {\rm det}(g + F)}\;  \left[    1  \; -\;
 {1\over 24}{(4\pi^2\alpha^\prime)^2 \over 32 \pi^2}\times \right.
 \cr
&
\times \left. {-1\over 16}\left( R^2_{(1)} + R^2_{(2)}
+ R^2_{(3)} -
R^2_{(4)} -{1\over 2} R^2_{(5)} -{1\over 2} R^2_{(6)} + R^2_{(7)} - R^2_{(8)} \right) +
\ldots  \ \right] }}

Given the way we have arrived at these corrections we cannot be
absolutely sure that there are no additional corrections of type $R^2$ that do not
contribute to the string amplitude. Indeed even for the $R^2$
corrections in \BachasUM\ there was an additional Gauss-Bonnet
term for which the coefficient could not be fixed by the string
amplitude and was fixed (to zero) by other means. Given that our
corrections reduce to the $R^2$ terms in \BachasUM\ in the limit
$F \to 0$, any additional corrections must vanish in this limit
and cannot contribute to the amplitude. We have tried many
generalisations other than \short\ and these requirements seem
hard to satisfy, but we haven't proven it is impossible.

There are other corrections to the D-brane action at order
$\alpha'^2$. In reference \WyllardYE\ corrections involving four
derivatives of $F$ where determined to all orders in $F$, using
boundary state formalism. This implicitly contains corrections of
type $\Omega^4$ where $\Omega$ is the second fundamental form, by
applying T-duality. To our knowledge, other possible terms such as $R \cdot
\Omega^2$ have not been determined to all orders in $F$.

\bigskip
\noindent
{\sl Acknowledgements:}

It is a pleasure to thank R.~Schiappa, and especially M.~Gutperle
and S.~Zhukov for useful discussions. This work was sponsored in
part by  grant NSF-PHY/98-02709.

\listrefs

\bye